\documentclass[onecolumn,authoryear]{els-mrw} 

\usepackage{amsmath,amssymb,amsfonts,amsthm,graphicx}
\usepackage{txfonts}
\usepackage{helvet}
\usepackage{xcolor}
\usepackage[normalem]{ulem}
\usepackage{hyperref}

\begin{document}

\chapter{A Primer on Dark Matter}\label{chap1}

\author[1]{Csaba Balazs}%
\author[2,6]{Torsten Bringmann}%
\author[3,4,6]{Felix Kahlhoefer}%
\author[5]{Martin White}%

\address[1]{\orgname{Monash University}, \orgdiv{School of Physics and Astronomy}, \orgaddress{Melbourne, Victoria 3800 Australia}}
\address[2]{
\orgname{University of Oslo}, \orgdiv{Department of Physics}, \orgaddress{N-0316 Oslo, Norway}}
\address[3]{\orgname{Karlsruhe Institute of Technology}, \orgdiv{Institute for Theoretical Particle Physics}, \orgaddress{76128
Karlsruhe, Germany}}
\address[4]{\orgname{Karlsruhe Institute of Technology}, \orgdiv{Institute for Astroparticle Physics}, \orgaddress{76128
Karlsruhe, Germany}}
\address[5]{\orgname{University of Adelaide}, \orgdiv{ARC Centre of Excellence for Dark Matter Particle Physics \& CSSM, Department of Physics}, \orgaddress{Adelaide SA 500 Australia}}

\address[6]{Corresponding authors: 
\email{torsten.bringmann@fys.uio.no}, 
\email{kahlhoefer@kit.edu} }

\maketitle

\begin{glossary}[Glossary]

\term{Annihilation} Process in which a particle-antiparticle pair disappears, converting its rest mass into energetic 
lighter particles. For example, an electron-positron pair can annihilate into a pair of photons.

\term{Antimatter} For all charged particles there exist so-called 
anti-particles with opposite charge but otherwise identical properties (like the mass). For example, the antiparticle of the electron is the positron. The present universe is known to contain much more matter than antimatter.

\term{Baryons} Composite particles that are made of three quarks, such as protons and neutrons. 
In astrophysics and cosmology, the term typically refers to all objects made of ordinary atomic
matter, which in particular also includes electrons.

\term{Bosons} Particles with integer spin. Bosons do not obey the Pauli exclusion principle, meaning that there is no 
limit to the number of identical particles in the same quantum state.

\term{Cross section} Measure of the probability of particles to interact with each other. The interaction rate, i.e.~the 
number of interactions per time, is obtained by multiplying the cross section with a particle flux (i.e.~number density 
times velocity).

\term{Electronvolt} Energy obtained by an electron as it crosses a potential difference of 1 volt. In particle physics this 
unit is also commonly used to measure rest mass (implicitly using $1\,{\rm eV}=1.78\cdot10^{-36}c^2$\,kg and 
setting the speed of light $c = 1$).

\term{Fermions} Particles with half-integer spin. Fermions obey the Pauli exclusion principle, meaning that no two 
identical particles may occupy the same quantum state.

\term{General relativity} Theory describing all forms of gravitational interaction. General relativity is often quoted as 
one of the best tested scientific theories, correctly reproducing observations
 from millimeter distances to solar system tests, gravitational waves and black holes.

\term{Phase space distribution} Distribution of particles in position and momentum space. The integral of the phase space distribution over all momenta yields the number density as a function of position.

\term{Power spectrum} Fourier transform of the autocorrelation function of a given cosmological observable. 
For example, the CMB temperature power spectrum describes how likely it is that a given sky patch has the same 
temperature, as a function of the angular size of this patch.

\term{Standard Model of Cosmology} (or $\Lambda$CDM)
The presently best global description of cosmological data in terms of 6 phenomenological parameters. Also referred to as the
`concordance model', it describes an expanding universe that is homogeneous and isotropic on large scales.

\term{Standard Model of Particle Physics} 
Theory describing the three known fundamental forces (excluding gravity) as well as all known elementary particles. Allows physicists to perform high-precision calculations, matching observations in particle-physics experiments.

\term{Virial theorem} Fundamental relation between the average kinetic and potential energy of a  
bound system of particles. A system is said to be virialized when it satisfies this relation,
meaning that the system has achieved a kind of statistical (but not thermodynamic) equilibrium.
\end{glossary}\\[-5ex]

\begin{glossary}[Nomenclature]
\begin{tabular}{@{}lp{34pc}@{}}
BBN &Big Bang Nucleosynthesis\\
CMB& Cosmic Microwave Background\\
$\Lambda$CDM & Lambda Cold Dark Matter (the cosmological concordance model)\\
\end{tabular}
\end{glossary}

\begin{abstract}[Abstract]

Dark matter is a fundamental  constituent of the universe, which is needed to explain a wide variety of astrophysical 
and cosmological observations. Although the existence of dark matter was first postulated nearly a century ago and its 
abundance is precisely measured, approximately five times larger than that of ordinary 
matter, its underlying identity remains a mystery. 
A leading hypothesis is that it is composed of new elementary particles,
which are predicted to exist in many extensions of the Standard Model of particle physics. 
In this article we review the basic evidence for dark matter and the 
role it plays in cosmology and astrophysics, and discuss experimental searches and potential candidates. Rather than 
targeting researchers in the field, we aim to provide an accessible and concise summary of the most important ideas 
and results, which can serve as a first entry point for advanced undergraduate students of physics or 
astronomy.
\end{abstract}

\newpage
\section{Introduction}

The term  `dark matter' is generally used to denote apparently missing mass contributions in 
various cosmological and astrophysical systems.
The name refers to its main defining properties: `Matter' in cosmology refers to objects that move with
non-relativistic velocity (much smaller than the speed of light) and cluster under the influence of gravity during the key 
stages of cosmological evolution --
in contrast to radiation, which has relativistic velocity and therefore possesses non-negligible pressure (counter-acting 
gravitational clustering). `Dark' 
implies that it does not emit or absorb light in the same way as visible matter, and therefore evades detection with 
conventional astrophysical methods. 

Historically, the term dark matter was used also to refer to certain forms of ordinary matter, such as very faint stars, 
planets or gas, which fell below the sensitivity of available telescopes but were at least in principle observable. 
Nowadays the term is used to refer to a hypothetical form of matter, which evolves dominantly under the influence of 
gravity -- and which thereby has an indirect but distinctly observable effect on astronomical objects such as stars or entire galaxies.
This is very different from ordinary atoms, molecules, ions and nuclei (collectively referred to as `baryonic matter' in 
cosmology), which are influenced by strong, weak and electromagnetic interactions and experience cooling, star 
formation, supernova explosions and many other effects.

It is important to emphasize that this definition of dark matter does not imply that all other types of 
interactions are absent. Instead, dark matter can participate in various additional processes, as long as these do not 
strongly affect the evolution on astrophysical and cosmological scales. Indeed, almost all dark matter models predict 
such additional processes, which may lead to potentially observable effects. This is what makes dark matter research 
exciting: we can hope to distinguish different dark matter candidates through their signatures in laboratory 
experiments or through subtle effects on astrophysical and cosmological observables. 
Hence, a broad program of experimental and observational efforts is underway to 
identify the fundamental nature of dark matter.
Understanding the nature of dark matter is key to solving fundamental questions about the structure and evolution of the universe.

In this article, we review the main pieces of evidence for dark matter, the key properties of dark matter established by 
observations and experiments and the most promising candidates that satisfy these requirements. The review is 
mostly aimed 
at advanced undergraduate students of physics or astronomy. While some familiarity with the basics of quantum mechanics, 
statistical physics and particle physics will be helpful, no prior knowledge of cosmology or quantum field theory is 
needed if the individual sections are read in the order presented here.  
In contrast to the excellent reviews 
by~\cite{Bertone:2016nfn,Bauer:2017qwy,Cirelli:2024ssz} and textbooks 
by~\cite{Profumo:2017hqp,Mambrini:2021cwd,Marsh:2024ury}, we restrict ourselves to the most important arguments and results in order to summarize the case for, and status of, dark matter as concisely as possible.

This article is organized as follows. We start, in section \ref{chap1:sec1}, with a brief introduction to
cosmology, and the crucial impact that dark matter has on the evolution of the universe
at the largest observable scales. Section \ref{chap2:subsec1} focuses on the impact of dark matter in the
present universe, on distance scales of galaxy clusters and below. In section \ref{sec:searches} we describe how 
non-gravitational interactions of dark matter particles may manifest themselves, and we provide an overview of 
the vast experimental program to  look for corresponding signatures. Finally, in section \ref{sec:models}, we briefly
present some of the most promising dark matter candidates, 
before concluding in section \ref{sec:summary}.

\section{Dark matter in the early universe}\label{chap1:sec1}

At the largest observable scales, the universe is isotropic, i.e.~it looks the same in all directions. Together with the additional
assumption that our Galaxy, the Milky Way, is not in any way located at a special position, this is often referred to as 
the \emph{cosmological principle}: it states that 
the universe as a whole is homogeneous and isotropic when {\it averaged} over sufficiently large regions of space. 
When applied to our leading theory of gravity, general relativity, this leads to the remarkable conclusion that
the universe is not only expanding presently (as directly confirmed by observations), but that it has been doing
so for at least 13\,billion years starting from an extremely hot and dense plasma. One of the most striking direct 
pieces of evidence for this {\it Big Bang theory}
is the existence of the cosmic microwave background (CMB), a highly isotropic leftover radiation that provides a 
snapshot of the state of the universe during these very early stages (see below).  

On the other hand, the description of the universe as being completely uniform is clearly very much at odds with our 
experience of an inhomogeneous world and an anisotropic sky. A central task for cosmology is therefore to explain the 
multitude of 
structures in the present universe while being consistent with the cosmological principle.
It is at this step, when explaining the existence and evolution of inhomogeneities, that dark matter appears as
an essential ingredient in our understanding of the cosmos.  
In contrast, the evolution of quantities that describe the universe as a whole -- like its age, expansion rate or
average density -- is insensitive to the microscopic properties and interactions of matter, and hence does not
require any distinction between dark and ordinary matter.

\subsection{A few basics of cosmological perturbation theory}\label{chap1:subsec1}

Let us start by introducing a bit of necessary notation. The expansion of the universe is described by the 
{\it scale factor} $a(t)$, which is a growing function of time $t$.
The scale factor essentially relates any so-called `comoving'  distance $d_{\rm com}$, namely the difference  between 
the coordinates of two objects, to a proper physical distance $d_{\rm phys} = a\cdot d_{\rm com}$ between these 
objects. Two galaxies without appreciable peculiar motion, for example, have a constant comoving distance 
$d_{\rm com}$, but their physical distances will still grow with time as $a(t)$, with a recession speed that is directly 
proportional to their distance.  In fact, this is exactly what Edwin
Hubble observed in the 1920s to be true for {\it all}
visible galaxies, and which let him conclude that the universe is expanding.

\newpage
The {\it average energy density} $\bar\rho$ of the universe can only depend on time, not space, because of the cosmological 
principle. This is of course not true for the local density $\rho$ at any given position $\mathbf{x}$. We denote this local
density as $\rho(t,\mathbf{x})\equiv \bar\rho (t)\left[  1+\delta(t,\mathbf{x}) \right]$, where we have introduced the 
{\it density contrast}  $\delta$. This density contrast thus indicates whether the local density is larger ($\delta>0$) 
or smaller ($\delta<0$) than the average cosmological density; in the former case we speak of a local {\it overdensity}.
How is such an overdensity expected to evolve in time? 
The two most important effects in this context are gravity and the intrinsic pressure of radiation: 
The former is always attractive and will hence tend to increase the overdensity,  
while the latter is repulsive and will therefore tend to decrease it. If an overdensity extends over a sufficiently large region of
space, the gravitational pull always dominates and the overdensity grows, attracting more and more matter from the 
surrounding regions (as long as this is efficiently possible, i.e.~until close to depletion).
This is known as the Jeans instability.

We can further refine the above description by introducing average densities $\bar\rho_i$ and density contrasts 
$\delta_i$ for each of the various main components that the universe consists of. Examples for this subscript $i$ 
include in particular any form of `radiation' ($i={\rm r}$), i.e.~massless particles such as photons or matter moving at 
velocities close to the speed of light, as well as more slowly moving or `cold' matter ($i={\rm m}$). 
At very early times, as evidenced by the CMB, the 
universe was highly homogeneous, which implies $|\delta_i|\ll1$ for these components. Thus, the underlying 
equations describing density contrasts become linear, since terms involving powers of $\delta_i$ are subdominant 
compared to $\delta_i$ itself. This greatly simplifies
the quantitative analysis  of how such tiny overdensities evolve in an expanding universe.
In fact, the resulting equations can now be solved fully analytically. One finds that
overdensities in radiation, $\delta_r$, do not 
grow and instead slowly decrease in time. Overdensities in matter, on the other hand, 
grow at the same rate as the universe expands:
\begin{equation}
\delta_m(t)\propto a(t)\,. 
\label{eq:at}
\end{equation}
Note that this crucial result is in accordance with our rough qualitative considerations above.
It assumes that there are no additional interactions between the different components, and that the 
overall energy density of the universe, $\bar\rho$, is dominated by rest mass -- as is the case during the times
relevant for our discussion here.
 As we will explain further down, the situation becomes more complicated once the perturbations start to become 
 non-linear, $\delta_m\gtrsim1$, and sophisticated numerical simulations have to replace the simple 
 analytic treatment sketched here.

\begin{figure}[t]
\centering
\includegraphics[width=.95\textwidth]{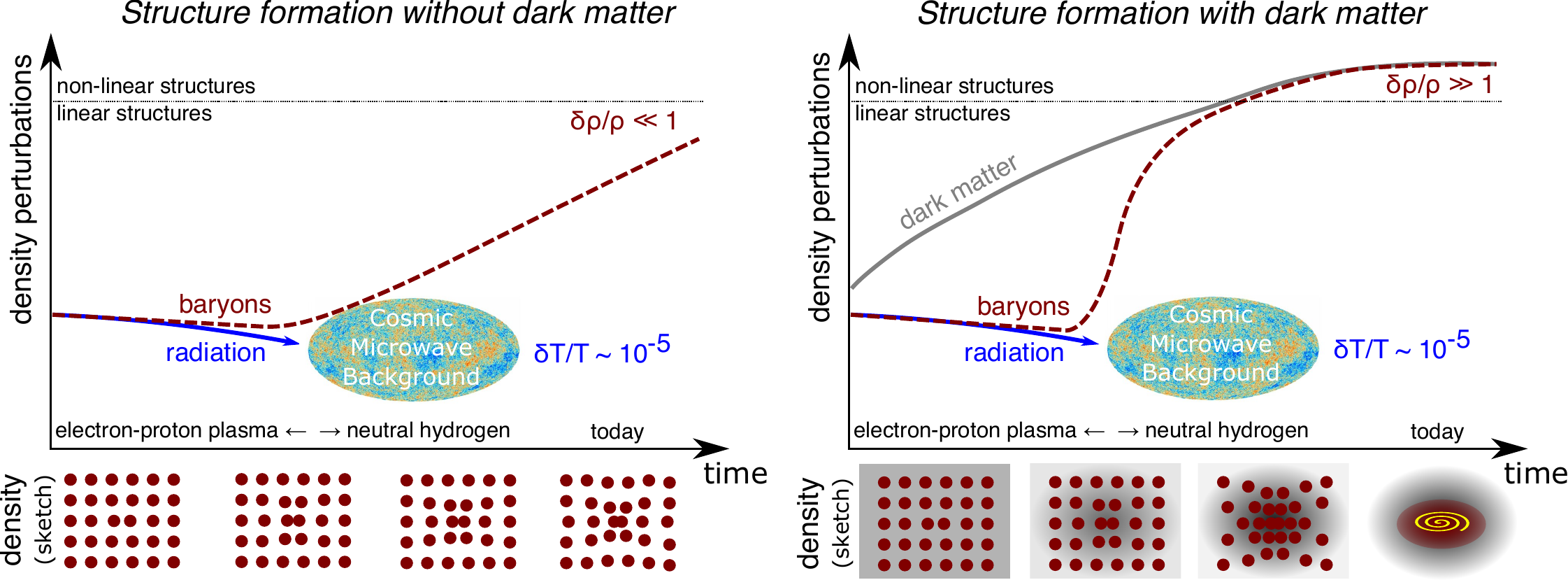}
\vspace{1.5mm}
\caption{
Illustration of the formation of structures in the early universe without (left) and with (right) dark matter. Only in 
the latter case do non-linear structures form that lead to the creation of stars and galaxies (yellow). In the former 
case, in contrast, the universe would remain largely uniform, developing only rather small density perturbations in 
ordinary matter (red). This largely agreed-upon picture rests on two observational inputs: {\it i)} the temperature of 
the nearly uniform CMB radiation, and {\it ii)} the tiny fluctuations of this temperature, depending on the  
direction in the sky (here strongly enhanced for visual purposes).}
\label{fig:structure_formation}
\end{figure}

\subsection{Evidence for dark matter}\label{chap1:subsec2}
The early universe was  a hot plasma of electrons and protons that frequently interacted with each other and with photons. 
As the universe expanded
and cooled down, electrons and protons combined to neutral hydrogen  -- releasing the photons that we now
observe as CMB. The observation of this background radiation, combined with the basic understanding of 
linear perturbation theory sketched above, immediately demonstrates that the universe cannot only consist of 
ordinary matter (commonly denoted as `baryons' in the cosmological context). Let us take this important 
argument step by step:
\smallskip
\begin{enumerate}
\item Initially, the released CMB photon energy corresponded to the plasma temperature $T_\text{CMB} = 0.3$\,eV:
at this point neutral hydrogen forms because there are no longer sufficiently many energetic photons 
to overcome the  ionization energy of hydrogen, $E_H\simeq 13.6\,$eV.
As the universe expanded the wavelengths of these photons were stretched, similar to the well-known Doppler effect 
for sound waves, thus resulting in a decrease of their frequencies and hence energies. 
The observed CMB temperature of $T_{\rm CMB}^{\rm today}\sim2.7\,{\rm K}\sim 0.2\,$meV  therefore implies that 
the universe must have been a factor of about $a_{\rm today}/a_{\rm CMB}\sim 10^3$ 
smaller at CMB times than it is today. 
\newpage
\item The CMB is not exactly isotropic, but shows tiny temperature variations $\Delta T/T\sim 10^{-5}$ across the sky.
This directly translates into corresponding density variations in the photon-baryon plasma just before 
neutral Hydrogen formed: $\delta_r(t_{\rm CMB})\simeq \delta_b(t_{\rm CMB})\sim 10^{-5}$.
\item After the release of the CMB photons, baryons come dominantly in the form of free neutral Hydrogen. 
Being non-relativistic (`matter'), their density perturbations should thus evolve as given by eq.~(\ref{eq:at}). In other
words, we should expect a density contrast in the present universe
of only $\delta_{\rm today}\sim a_{\rm today}/a_{\rm CMB} \cdot \delta_b(t_{\rm CMB})\sim 10^{-2}\ll1$.
\end{enumerate}
\smallskip
This describes a universe that is largely homogeneous -- with {\it no} structures such as stars or galaxies -- and hence
very far from what we observe. The only way to circumvent the above chain of arguments is by 
somehow boosting the growth of $\delta_b$, from the initial value of $10^{-5}$, to proceed significantly faster
than $a(t)$. This requires the existence of sizable gravitational potentials already at CMB times,
which the hydrogen would be attracted by and fall into. Fig.~\ref{fig:structure_formation} further 
illustrates this point.
Such gravitational wells cannot easily be maintained in a tightly coupled plasma.
They  require a significant  contribution to the overall energy density that is much less interacting than any form of 
ordinary, visible matter -- and which we therefore denote as {\it dark matter}.

\begin{figure}
\begin{BoxTypeA}[chap1:box1]{\normalsize\textsf{The Cosmic Microwave Background: A closer look for the interested reader}}

\vspace{2mm}
\centering
\begin{minipage}{0.5\textwidth}
The arguments given in the main text, and illustrated in figure \ref{fig:structure_formation}, can be significantly 
refined. The CMB anisotropies, in particular, 
have a very characteristic scale dependence: deviations from the mean CMB temperature  
are by far most commonly observed 
within patches of the sky with an angular size of about $1^\circ$, followed by 
corresponding `blobs' of roughly $0.3^\circ$ and $0.2^\circ$ in size. 
Mathematically, this is elegantly described in terms of the Fourier transformation of $\Delta T/T (\psi)$,
with $\psi$ denoting the direction in the sky. This Fourier transform is known as the {\it power spectrum} of CMB 
temperature fluctuations. Scales on which temperature fluctuations are more likely appear as peaks in the power 
spectrum. As stressed in the short introduction to
cosmological perturbation theory, the evolution of small (`linear') density perturbations can be studied with almost 
arbitrary theoretical precision, for all energy and matter components individually.
This allows scientists to compute the theoretically expected form of the power spectrum as a function of 
the 6 $\Lambda$CDM parameters.
\smallskip

For a suitable choice of these parameters, the result perfectly matches the observed power spectrum -- which is 
a highly non-trivial consistency check of the model. In this way, the 6 $\Lambda$CDM parameters can in fact 
directly be `measured', with percent-level accuracy. 
Two of these parameters are of particular importance in our context: the 
average cosmological energy densities in baryons and dark matter, respectively. The ratio of these densities affects
 the relative
height of the even and odd peaks in the power spectrum, which in turn reflect the oscillations in the primordial 
plasma due to the competition between the attractive force of gravity (mostly due to dark matter) and 
the repulsive force of the radiation pressure (preventing the plasma with ionized baryons to be compressed too 
densely). It is particularly the ratio of the various peak heights that is notoriously
difficult to correctly account for in alternative explanations of the CMB anisotropies.

A third parameter of the $\Lambda$CDM model, the present-day expansion rate, is related to the total energy density 
of the present universe. It implies that the dominant form of energy today is in fact neither baryonic nor dark matter, 
but a mysterious energy form called dark energy, which we will not discuss further here. 
\end{minipage}
\begin{minipage}{0.05\textwidth}
\quad
\end{minipage}
\begin{minipage}{0.4\textwidth}
\begin{center}
\includegraphics[width=0.95\textwidth]{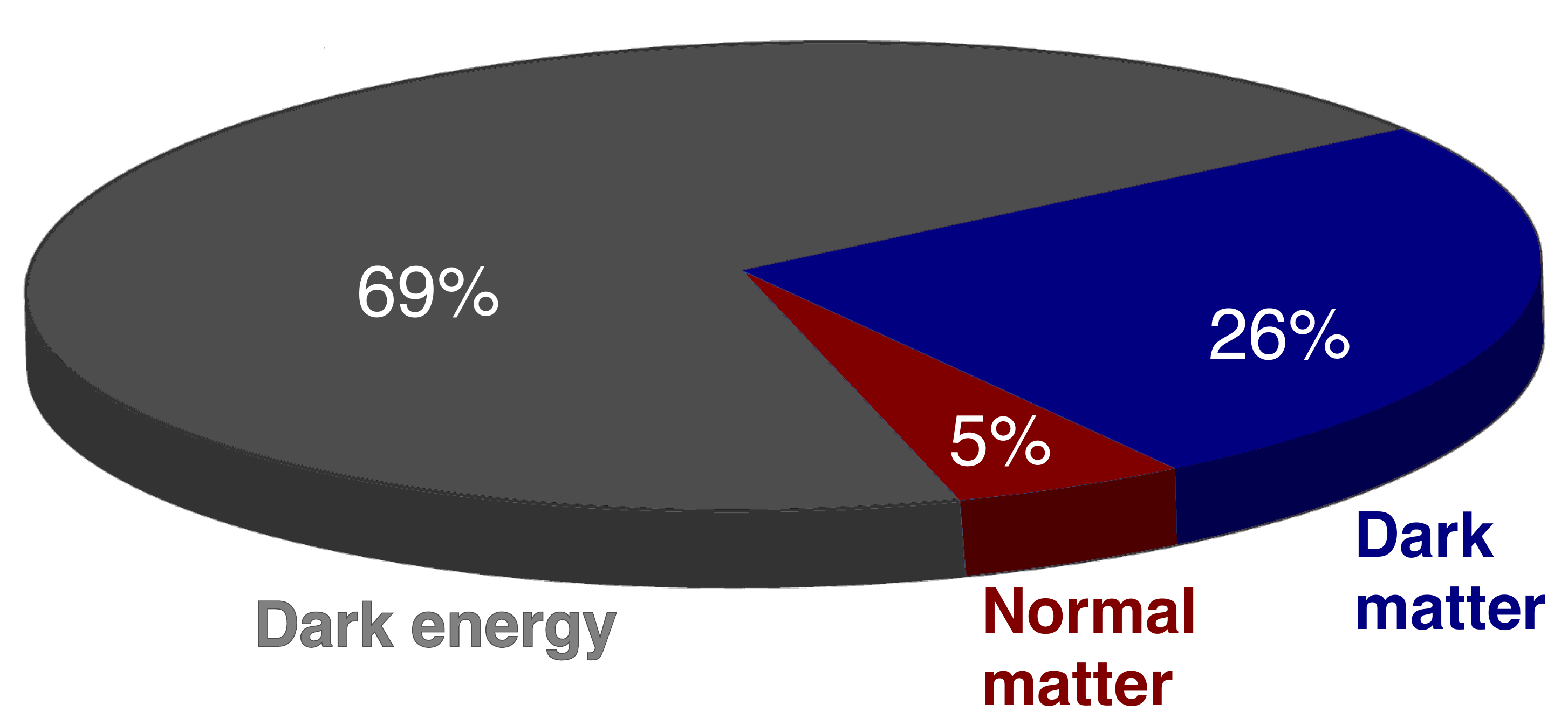}

\vspace{3mm}

\includegraphics[width=0.99\textwidth]{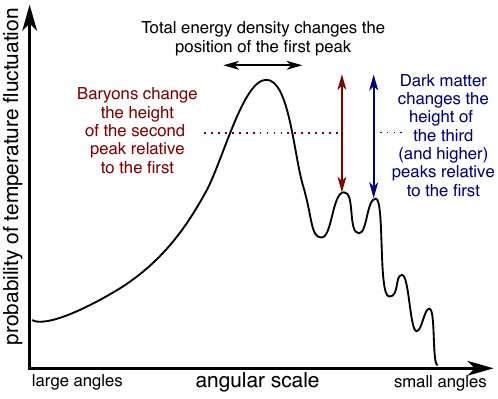}
\vspace{2mm}
\end{center}
\textsf{Top: Composition of the present universe as inferred from CMB data. Bottom: Illustration of the effects of baryons and dark matter on the CMB temperature power spectrum.}

\end{minipage}
\end{BoxTypeA}
\end{figure}

As time passes, the initially tiny density fluctuations in both the ordinary and dark matter components grow under the 
effect of gravity. Density contrasts in the latter are the first to reach sufficiently large values for overdense regions to 
gravitationally collapse, forming a network of extended filaments 
and at their intersections bound systems that quickly virialize 
-- thus re-distributing potential and kinetic energy -- to dark matter {\it halos}. 
Eventually, visible galaxies composed of
ordinary matter then form at the center of these dark matter halos, as explained in section \ref{chap2:subsec1}. 
The distribution of ordinary matter thus provides a useful tracer of the dark matter distribution, in particular
at the largest scales, where the average density contrasts have only become mildly non-linear. 
At these scales, indeed, one observes the imprint of the same primordial plasma fluctuations that we 
discussed above for the CMB -- just this time directly in the distribution of ordinary matter, so-called baryon acoustic 
oscillations. 
It is possible to describe all the different observations with a single model with only six parameters,
known as the Standard Model of Cosmology, or $\Lambda$CDM in short.

The field of cosmology has turned into a true precision science within the last few 
decades, with several Nobel prizes awarded to fundamental cosmological observations and their
theoretical understanding, not the least related to the role of structure formation.
Dark matter is an absolutely essential building block of what today is
referred to as the concordance model of cosmology. At this point, in other words, it has become almost misleading
to refer to  the various cosmological observations as merely providing evidence for dark matter; rather, they essentially
constitute precision {\it measurements} of its properties. Below, we summarize some of the most important 
conclusions.

\newpage
\subsection{Implications for dark matter models}

The plethora of cosmological observations cursorily 
mentioned above do not only precisely 
tell us how much dark matter there is in total, they also put stringent constraints  on its properties. 
In particular, dark matter 
must be fundamentally different from ordinary matter in order to avoid severe problems with cosmological
structure formation. This is consistent with our current understanding of big bang nucleosynthesis (BBN), which 
explains how light elements like hydrogen, helium and lithium were formed during the first seconds of the universe
-- well before the CMB.
The successful predictions of the cosmologically observed abundances of these elements imply,
independently, that the total amount of ordinary matter that was present during BBN
falls short by a factor of about five with respect to the total amount of matter. 
Both BBN and CMB considerations also leave very little room for any form of energy conversion between these components, 
i.e.~any mechanism that would dump energy from dark matter into the primordial plasma of ordinary matter
(see also section \ref{sec:searches} for further constraints on how dark matter might interact with ordinary matter).

Remarkably, cosmological considerations even put tight constraints on dark matter models that do {\it not} interact
with ordinary matter in any appreciable way. For example, dark matter must be stable on cosmological timescales 
in order to be consistent with structure formation; quantitatively, at most a few percent of the dark matter can have 
decayed since CMB times -- independently of whether the decay products include Standard Model particles or not. Furthermore, dark matter moving at slightly too large velocities after CMB times 
-- coined `warm' dark matter -- would escape from overdense regions before they fully collapse, 
preventing the formation of small gravitationally bound objects such as dwarf galaxies.
For thermally produced dark matter (see section \ref{sec:models}), for example, this excludes particles with masses
lighter than a few keV. 
At much earlier times, dark matter can, and in general will,  move much faster. However, this
must not contribute more than about $\mathcal{O}(10\%)$ to the relativistic 
energy content of the universe at the times of BBN, in order 
to not to alter the expansion rate and thereby 
spoil the successful prediction of the light element abundances. 
Using again thermally produced dark matter 
as an example, this translates to a lower mass bound of a few MeV.

Notably, these constraints also apply to dark matter models that do not involve new elementary particles. 
For example, primordial black 
holes as putative dark matter candidates (see section \ref{sec:models}) 
cannot be unstable, which would be the case for masses below 
about $10^{15}$\,g, due to Hawking radiation. In addition, they must not convert too much of their rest mass into 
gravitational waves during merger events 
(which constitute a form of radiation). There exist also interesting attempts to modify 
general relativity as our currently established description of gravity to avoid postulating the existence of dark matter at all, but all of these attempts currently face severe
challenges in addressing the multitude of cosmological observations and constraints only briefly touched upon here.
\vspace{-1mm}

\section{Dark matter in the present universe}\label{chap2:subsec1}
\vspace{-0.5mm}
In the previous section we have discussed how the growing dark matter density perturbations create the seeds for the 
formation of visible structures. The baryonic gas trapped in the overdensities cools down and forms clouds of 
molecular hydrogen. Eventually, sufficiently large densities and temperatures are reached to reionize 
the hydrogen gas into free electrons and protons, which leads to a further contraction of the gas and the onset of 
nuclear fusion, i.e. the formation of the first stars. These structures are created at all length scales, but small 
structures (individual galaxies) form first and subsequently merge to form larger structures (such as galaxy clusters). 
The implication of this entire process is clear: the distributions of dark and visible matter in the universe are 
necessarily strongly correlated. 

\begin{figure}[t]
\centering
\includegraphics[width=.95\textwidth]{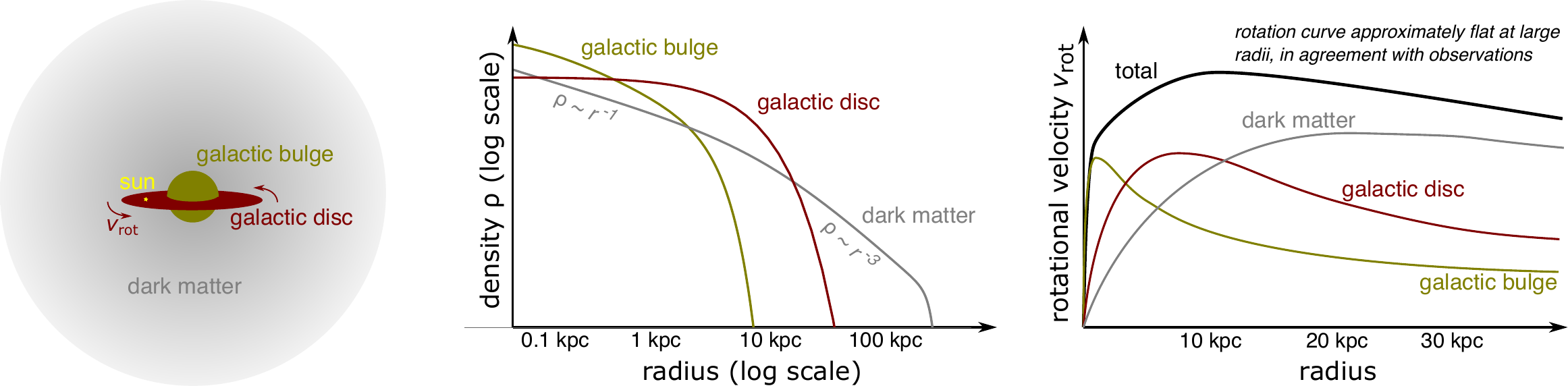}
\vspace{1.5mm}
\caption{Left: Illustration of a spiral galaxy like the Milky Way embedded in a typical dark matter halo. 
Our own star, the Sun, is located at a distance of about 8.5\,kpc from the center of the Milky Way.
Center: Density profiles of the different components in the plane of the disk, as a function of distance 
from the center. 
For the dark matter component we choose the Navarro-Frenk-White profile, eq.~(\ref{eq:nfw}), up to the 
edge of the halo.
Right: Contribution of the different components to the gravitational potential that determines the rotational velocity.
Note that the disk and bulge distributions can vary significantly between differently galaxies, which in turn
also affects the theoretically expected dark matter density profile.}
\label{fig:halo}
\end{figure}

Indeed, large-scale numerical simulations of structure formation that include the effects of both gravity and the 
hydrodynamics of the baryonic gas predict that galaxies and galaxy clusters typically 
sit in the centres of so-called dark matter 
halos, self-gravitating systems of dark matter. These systems are expected to be virialised, i.e.~to have an 
approximately
time-independent phase space distribution that obeys the \emph{virial theorem}. In contrast to the baryonic gas, which 
can cool and collapse into disk-like structures, dark matter halos cannot dissipate energy via the emission of photons 
and therefore remain more diffuse and three-dimensional (see figure~\ref{fig:halo}). 

In fact, most dark matter halos in numerical simulations of structure formation are nearly spherically symmetric and 
have radial density profiles that can be described by a small number of parameters. The simplest such density profile 
is the so-called isothermal halo, for which $\rho(r) \propto 1/r^2$. More detailed studies have shown that the density 
profiles of typical dark matter haloes fall more steeply at large radii, $\rho(r) \propto 1/r^3$, and less steeply in the 
inner regions, $\rho(r) \propto 1/r$. This leads to  the so-called Navarro-Frenk-White profile,
\begin{align}
\label{eq:nfw}
\rho(r) = \rho_s \frac{r_s}{r} \left(\frac{2}{1+r/r_s} \right)^2\,,
\end{align}
which is parametrized by a normalization constant $\rho_s$ and the scale radius $r_s$, where the slope of the density 
profile changes. While haloes in dark matter-only simulations more or less universally follow this form,
state-of-the-art simulations that include the complicated physics associated with ordinary matter show a certain diversity
in the density profiles of dark matter haloes, in particular for smaller haloes.
But the most important prediction of numerical simulations of structure formation is a much simpler one: if dark 
matter surrounds most visible structures, it should be possible to  identify the effect of 
its gravitational interactions by closely observing galaxies and galaxy clusters.

The task is therefore to map out the gravitational potential of a given astrophysical system and compare it with 
expectations based on the observed amount of visible matter. Broadly speaking, this can be done in two ways. The 
first is to obtain kinematic data from the visible constituents of the system, i.e.~stars and gas, and infer the 
gravitational 
acceleration. The more energetic the visible matter, the larger must be the gravitational attraction to keep the system 
stable. The second is to use gravitational lensing, which is the effect that a gravitating mass has on the curvature of 
space-time and hence on the propagation of light. Indeed, any astrophysical system can act as a gravitational lens for 
background objects. Depending on the mass and distance of the lens, one may observe multiple images, distortions 
or a time-varying brightness of the lensed object -- which can be used to reconstruct the  gravitational potential
and hence mass distribution.

\subsection{Dark matter in galaxy clusters}

\begin{figure}[t]
\centering
\includegraphics[width=.7\textwidth]{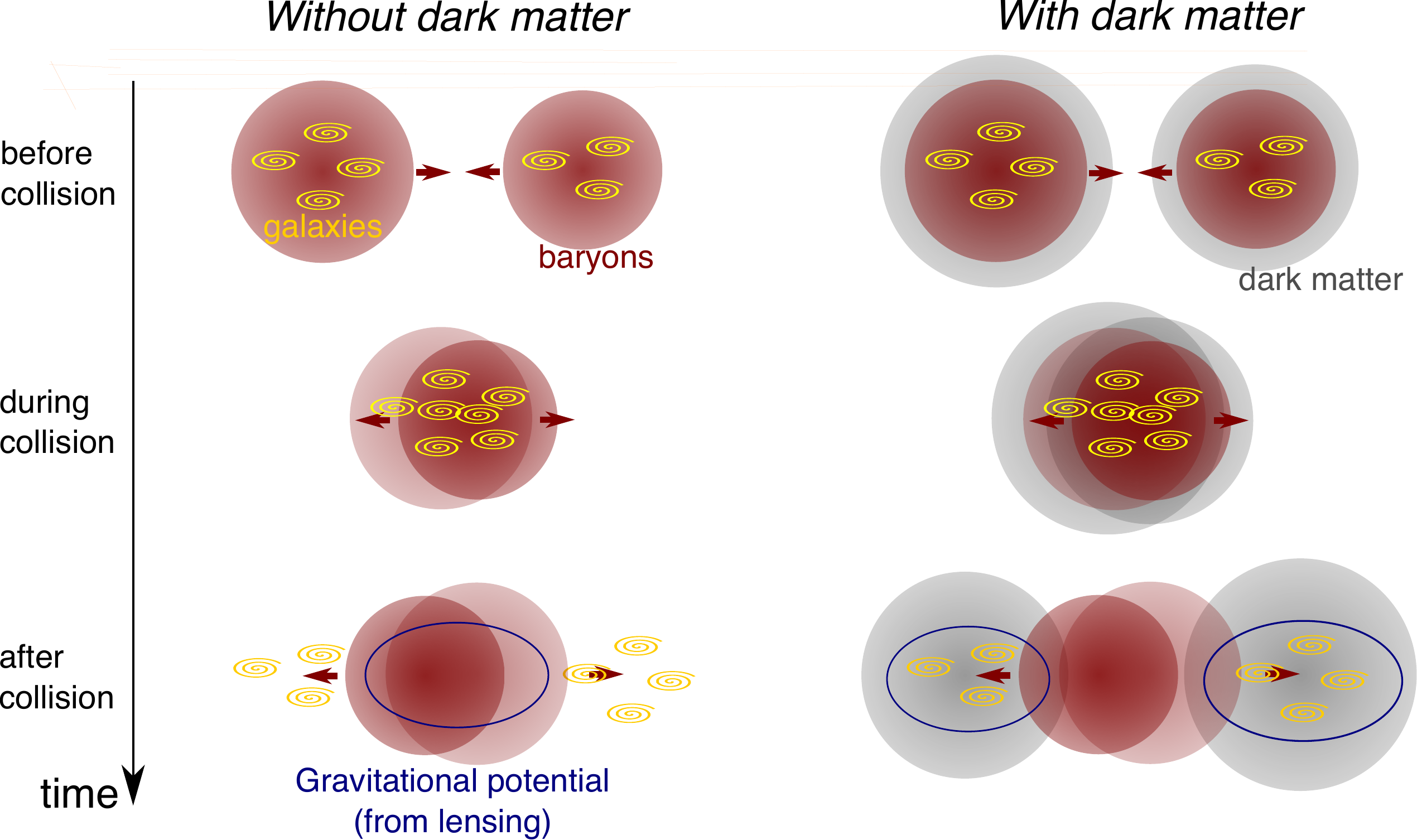}
\vspace{1mm}
\caption{Illustration of the collision of two galaxy clusters without (left) and with (right) dark matter. In the absence of 
dark matter (grey), the baryonic gas (red) constitutes the dominant form of matter in a galaxy cluster. During the collision it 
experiences friction and heats up, which causes it to slow down relative to the individual galaxies, such that the 
gravitational potential inferred from gravitational lensing is naively expected to peak in the centre 
between the two clusters. If dark matter constitutes the dominant form of matter in galaxy clusters, instead, 
the gravitational potential after the cluster collision exhibits two clearly separated peaks that coincide with the galaxies.
}
\label{fig:BC}
\end{figure}

The most impressive piece of evidence for dark matter stems from the observation of colliding galaxy clusters, i.e.~galaxy clusters 
that are pulled towards each other by gravity until they collide and eventually merge. In the process, the baryonic gas, 
which constitutes the dominant form of visible matter in galaxy clusters, is slowed down by friction and 
becomes spatially separated from the  stellar component, which is to a very good approximation collisionless. This 
separation can be easily observed given the X-ray emission of the hot baryonic gas. The trick is now to simultaneously 
measure the gravitational potential of such a system using gravitational lensing. In the absence of dark matter, one 
would expect the gravitational potential to be dominated by the baryonic gas and hence to coincide with the X-ray 
emission. However, observations clearly show that the gravitational potential peaks at the position of the stellar 
component, which can only be explained if there is an additional invisible and collisionless mass component, matching 
exactly the properties of dark matter. 
This is illustrated in figure~\ref{fig:BC}, which shows on the left the expected behaviour of colliding galaxy clusters in 
the absence of dark matter, and on the right the predictions in its presence. Observations of cluster collisions clearly 
agree with the latter case, most famously the so-called `Bullet' cluster 1E0657-558.

Evidence for dark matter can also be obtained from virialized galaxy clusters.
By measuring the temperature of the ionized gas, for example, one can infer 
the gravitational force that is needed to confine the gas to the galaxy cluster. This 
force is much larger than what can be provided by the gas itself, thus reinforcing the need for 
dark matter. Moreover, the required amount and distribution of dark matter quantitatively agrees with predictions from 
the CMB and numerical simulations of structure formation. In fact, one finds that typical galaxy clusters mirror 
the composition of the universe, in the sense that they contain about five times more dark than visible matter. Similar 
conclusions are reached when measuring the velocity dispersion of galaxy clusters and using the virial theorem to 
infer the gravitational potential. In fact, this strategy was already employed in the 1930s to obtain some of the 
historically first pieces of evidence for dark matter.

\subsection{Dark matter on galactic scales}

Smaller astrophysical systems also exhibit clear deviations of observations from the expectations based on the visible 
matter and Newtonian dynamics. The most famous and historically most important example are so-called galactic 
rotation curves, which refer to measurements of the rotational velocity of stars and gas in spiral galaxies as a 
function of their distance from the centre. For a spherically symmetric mass distribution, these velocities should scale 
as  $v_\text{rot}(r) \propto \sqrt{ M(r) / r}$, where  $M(r)$ denotes the total mass within 
radius $r$. If most of the mass of a galaxy is enclosed within some radius $R$, we hence expect the rotational velocity 
to exhibit a Keplerian decline proportional to $1/\sqrt{r}$ beyond this radius. Observations, however, show that 
$v_\text{rot}(r)$ remains approximately constant far beyond the edge of the visible disk, indicating that $M(r)$ grows 
proportional to $r$ even in regions of space where almost no visible matter is observed. These flat rotation curves 
therefore indicate the presence of a three-dimensional dark matter halo with a density profile that decreases 
approximately as $\rho(r) \propto r^{-2}$ 
on these scales, in accordance with what we discussed in the context of eq.~(\ref{eq:nfw}).
Intuitively, 
the gravitational attraction of dark matter is needed to prevent spiral galaxies from flying apart due to the large 
rotational velocities. 

Similar observations can also be made in even smaller systems, so-called dwarf spheroidal galaxies, in particular 
those that orbit the Milky Way as satellites. Using stellar kinematics, one can show that in these systems dark matter 
can dominate over visible matter by several orders of magnitude. As a result, these systems can be very faint, 
sometimes including fewer than a thousand stars. This makes it challenging to infer the dark matter density profiles in 
detail.

Let us briefly pause our discussion to mention that the evidence for dark matter using kinematic data is not 
uncontested. The reason is that these arguments assume the validity of Newtonian dynamics, i.e.~the proportionality 
of force and acceleration, $F \propto a$. However, the required accelerations are so tiny that the validity of Newtonian 
dynamics cannot be established in the laboratory, and hence it is in principle conceivable that the relation between 
force and acceleration is modified. Indeed, the observation of flat galactic rotation curves can also be explained if 
$F \propto a^2/a_0$ for $a < a_0$, where $a_0$ denotes a new fundamental constant of nature. While successful on 
galactic scales, this so-called Modified Newtonian Dynamics does not remove the need for dark matter on galaxy 
cluster and cosmological scales. For the moment, the dark matter hypothesis therefore remains the only viable 
explanation of all the observations discussed above.

\subsection{Dark matter in the Milky Way}

In spite of this wealth of evidence for the presence of dark matter in various astrophysical systems, determining the 
distribution of dark matter in the Milky Way is surprisingly difficult. This is because it is much harder to measure a 
rotation curve from the “inside” than for a galaxy seen from the side. Nevertheless, it is possible to fit an assumed dark 
matter density profile to data of stellar kinematics and use the result to infer the average 
dark matter density in the solar neighbourhood. 
An alternative approach that does not require the assumption of a specific density profile is to study the motion of 
stars in the direction perpendicular to the galactic disk. Both approaches assume that the dark matter distribution is 
smooth, i.e.~it does not vary significantly over the distance between nearby stars. In particular, the dark matter density 
within the solar system is assumed to be a time-independent constant, called the local dark matter density $\rho_0$.
Both methods give similar results and lead to a value of approximately 
$\rho_0 \approx 0.4\, \mathrm{GeV}\, \mathrm{cm^{-3}}  \approx 0.01 \, M_\text{sun} \, \mathrm{pc^{-3}}$, 
where the uncertainties can be as large as a factor of 2. It is worth emphasizing that this density is tiny -- more than 20 
orders of magnitude smaller than the density of visible matter in the Earth's atmosphere. In other words, dark matter is 
extremely dilute 
and its gravitational effects only become relevant on the vast distance scales relevant for astrophysics and cosmology.

When considering possible strategies to detect dark matter from the solar neighbourhood with laboratory experiments, we are also interested in the dark matter velocity distribution. Motivated by both analytical arguments and numerical simulations of structure formation, this distribution is usually modelled by a Maxwell-Boltzmann distribution
\begin{equation}
    f(v) \propto v^2 \exp \left( - \frac{v^2}{v_0^2}\right) \, ,
\end{equation}
where the velocity dispersion $v_0 \approx 240 \,\mathrm{km \, s^{-1}}$ is related to the rotational velocity of the Galactic disk 
inferred from stellar kinematics.
This distribution implicitly assumes an isothermal dark matter halo with $\rho(r) \propto r^{-2}$ and is therefore overly 
simplistic. In particular, the exponential tail of the distribution must be cut off, since the velocity of a virialised dark 
matter component cannot exceed the galactic escape velocity of approximately $600 \, \mathrm{km \, s^{-1}}$. 
However, there may well be non-virialized components, such as dark matter streams, with even higher velocities.
In general, the exact form of the velocity distribution will depend  on the details of the total density profile
and, to some extent, the recent merger history of the halo. 

\subsection{Implications for dark matter models}

To summarize this section, let us emphasize that astrophysical systems do not only provide further evidence for the 
existence of dark matter, but also further constrain its properties. In particular, it is clear from these observations that 
dark matter behaves fundamentally differently from baryonic matter, in the sense that it does not experience the same 
amount of collisions and dissipation. In fact, we can use the observations of the Bullet Cluster to derive an upper 
bound on the dark matter self-scattering cross section of 
$\sigma_\text{self} / m_\text{DM} \lesssim 2\, \mathrm{cm^2 \, g^{-1}}$. 
This is a rather weak bound, which is easily satisfied for example by Standard Model neutrinos. 
Nevertheless, self-scattering cross sections slightly below this bound may have interesting implications for structure 
formation, for example because they can modify the density profiles of dark matter halos in potentially 
detectable ways.

Another important puzzle piece comes from the fact that a self-gravitating gas of fermions must obey the Pauli 
exclusion principle, which constrains their phase space distribution. Namely, if a gas of fermions is confined in 
space, it must occupy higher momentum states. By measuring the mass and spatial extent of a dark matter halo, we 
therefore obtain a lower bound on the mass of the individual dark matter particles, which relates velocity and 
momentum. This is known as the Tremaine-Gunn bound. It is strongest for small systems such as dwarf spheroidal 
galaxies, giving a lower bound of approximately $m_\text{DM} \gtrsim 500 \, \mathrm{eV}$. While weaker than the 
warm dark matter bound discussed above, it is independent of the cosmological history and the evolution of 
structures. Thus, the mere existence of dwarf spheroidal galaxies excludes the possibility that dark matter is 
dominantly in the form of neutrinos, which are known to have much smaller masses. 

While the Tremaine-Gunn bound does not apply to bosonic dark matter particles, their masses are also constrained 
by dwarf spheroidal galaxies. The reason is that for extremely small momenta (i.e.~tiny dark matter masses) the 
Heisenberg uncertainty principle forbids the localisation of dark matter particles on the scale of these astrophysical 
structures, or below. Put differently, the de Broglie wavelength of such particles becomes larger than the size of dwarf 
spheroidal galaxies, incompatible with observations. The existence of such galaxies therefore gives a lower bound of 
$m_\text{DM} \gtrsim 10^{-21} \, \mathrm{eV}$ 
for bosonic particles. 
\vspace{-1mm}

\section{Experimental searches for dark matter}
\label{sec:searches}
\vspace{-0.5mm}

So far, we have reviewed the abundant evidence that dark matter exists and that we can in fact quantify the overall
amount of dark matter with percent-level accuracy. In this sense we have already “detected” dark matter. However, 
each piece of evidence has only involved the force of gravity and is compatible with the absence of any 
further interactions of dark matter. A moment’s reflection tells us that this is unsatisfactory, since it leaves us with no 
clear idea of what dark matter actually is.

To see this, imagine that we only knew of the gravitational interactions of visible matter, i.e.~the sorts of structures that 
visible matter forms in the universe. We would be ignorant of the laws of fundamental physics and chemistry, which 
ultimately lead on to biology and everything we are used to dealing with in everyday life! What we actually know is that 
the matter in the universe consists of a relatively small set of fundamental particles: the up, down, strange, charm, 
bottom and top \emph{quarks}, the electron, muon and tau \emph{leptons}, and the electron, muon and tau 
\emph{neutrinos}. 
Each particle (possibly with the exception of neutrinos) is accompanied by an anti-particle with opposite charge 
(e.g.~the positron for the electron), although the present universe is known to contain much more matter than antimatter.
The  Standard Model describes the interaction of these particles via three forces, named the 
\emph{strong}, \emph{weak} and \emph{electromagnetic} forces. Each of these is associated with extra particles that 
mediate the interactions, namely the photon (electromagnetism), W and Z bosons (weak force) and gluons (strong 
force). Finally, the Higgs boson explains why the quarks and leptons, as well as the W and Z bosons are massive.

The lesson from the visible sector is that our idea of “what something is”  requires a theory of fundamental objects and 
their interactions beyond gravity. The obvious way forward for the dark matter problem is therefore to design and 
perform experiments that probe the non-gravitational interactions of dark matter, which will ultimately tell us its 
fundamental nature. The Standard Model of Particle Physics even tells us how to start this process, since we can 
examine the known particles and ask how we would observe possible interactions between them and dark 
matter. There are several different ways that we might observe such interactions, and we will step through each in 
turn.

\subsection{Dark matter annihilations and decays}

Our understanding of dark matter halos indicates that there are large concentrations of dark matter particles within 
galaxies, galaxy clusters and dwarf satellite galaxies. Looking at such regions with our rich variety of 21st century 
astronomical instruments is therefore an excellent way to search for non-gravitational dark matter interactions.

The fundamental principles of particle physics suggest two ways that dark matter might produce visible matter. The 
first is that it might decay to produce Standard Model particles, just like a neutron decays into a proton, an electron 
and an anti-neutrino. The decay would have to be very slow indeed, so that most of the dark matter inferred from the 
CMB is still present today, producing the gravitational lensing and galactic kinematic signatures described earlier.
A second option is that dark matter might annihilate with dark antimatter\footnote{%
{We note that it is unknown whether dark matter and dark antimatter are equally abundant in the universe, or whether 
there is an asymmetry similar to visible matter. However, since dark matter particles do not carry electric charge, they 
could in principle be their own anti-particles, in which case there would be no distinction between dark matter and dark 
antimatter and any pair of dark matter particles could annihilate.}
} 
to produce visible particles, analogous 
to the annihilation of an electron-positron pair into photons. Such a process would naturally occur in regions of our 
universe where the dark matter (and dark antimatter) densities are at their highest. The challenge in searching for the 
annihilation products is that regular astrophysics processes themselves often create the same particles, and we must 
therefore look for regions of the universe where any dark matter signal is strong enough to be unambiguously 
observed above the background of regular astrophysics. Promising targets for observation are the centre of our own 
Milky Way galaxy (the nearest place with a huge amount of dark matter in it), dwarf galaxies (which have a low 
background from conventional astrophysics), or even the centre of our own Sun (which could have steadily 
accumulated dark matter for billions of years and holds it gravitationally-bound in its core). 

What sort of particles could be produced by annihilating or decaying dark matter? In principle, one should consider all 
of the particles in the Standard Model, but many of these particles decay almost immediately and thus cannot be 
detected on Earth. We therefore concentrate on particles that are stable on astrophysical distance scales. 

\newpage
{\bf Photons: } 
The `dark' in `dark matter' means that there can be no significant interaction between dark matter 
and photons, but the rules of the Standard Model, and of quantum mechanics more generally, mean that we cannot 
discount all possible ways to produce photons from dark matter. For example, dark matter might annihilate with dark antimatter to 
produce particles such as neutral pions that decay to photons, or particles such as leptons that emit photons via their 
charge. A charged electron moving on a curved trajectory through a galactic magnetic field, for example, will emit 
synchrotron radiation in the form of photons with a typical wavelength in the radio band. 
Even the direct annihilation into a pair of photons is possible, 
resulting in a signal that is very sharply peaked in the photon energy.
While necessarily highly suppressed, such a signal is still phenomenologically very important because it 
would clearly stand out from typical astrophysical backgrounds.
The same mechanisms that produce photons from dark matter 
annihilation are also relevant for dark matter decay.

The highest-energy photons produced by dark matter can be searched for with gamma-ray telescopes, which are 
either based in space (e.g.~Fermi-LAT), or operate on the ground by observing the air showers produced when high-energy photons hit the upper atmosphere (e.g.~H.E.S.S.~and the forthcoming Cherenkov Telescope Array). Other 
signatures may appear in the radio or X-ray bands, depending on the nature and interactions of dark matter. Such 
searches are highly sensitive despite the presence of significant astrophysical backgrounds. For example, for very 
heavy dark matter, null observations constrain the decay lifetime to be in excess of a billion trillion years.

{\bf Neutrinos: } Dark matter could produce neutrinos, either directly or by decaying to other particles that themselves 
produce neutrinos (e.g.~tau leptons). Although one could in principle search for an anomalous flux of neutrinos from 
the galactic centre and other distant sources, such searches tend not to be competitive with the search for other 
particles. 
Instead, an intriguing option is to look for a flux of neutrinos from our own Sun. The Sun has been passing through the 
dark matter halo of the Milky Way for billions of years. Interactions between dark matter and the nuclei in the Sun can 
cause dark matter particles to be captured, lose energy through further scattering processes and eventually sink to the solar core. 
If these particles annihilate to produces photons, these would never escape the Sun to be observed. Neutrinos, however, are so weakly 
interacting that they could reach Earth and be detected by experiments such as IceCube, which observes interactions between high-energy neutrinos and the ice of the South Pole. 

{\bf Electrons/positrons/antiprotons: }Other potential by-products of dark matter annihilation and decay include 
electrons, protons, positrons and antiprotons. These can travel to Earth and be detected by space-based cosmic ray observatories such as the AMS-02 experiment. Since by far most of the visible matter in space is 
comprised of protons and electrons, the contribution from dark matter processes is dwarfed by normal processes. 
However, antiprotons and positrons are much rarer, and any excess of these particles could offer our first glimpse of 
non-gravitational dark matter interactions. 
Excesses of both positrons and antiprotons over those expected from the cosmic ray-induced background 
have in the past been identified, but none of them could so far unequivocally be attributed to a dark matter-related
origin.

\subsection{Dark matter scattering and absorption}

Apart from annihilation, DM can also interact through scattering with nuclei.
This gives rise to an entirely different way of hunting for non-gravitational interactions of dark matter.
The relative motion 
between the Earth and the dark matter halo means that, if dark matter is comprised of particles, they constantly 
pass through the Earth. Each second, approximately 
$10^{-16}$ kg of dark 
matter (an amount equivalent to the mass of $10^{11}$ protons) passes through each of our bodies. 
By placing a low-background detector filled with suitable nuclei underground, we can look for 
dark matter particles that scatter on the nuclei and thereby transfer some of their kinetic energy, leading to a so-called `nuclear recoil'. 
Similarly, we can look for the scattering of dark matter off the electrons bound to the nuclei in the detector. 
Yet another 
way that dark matter can influence the detectors is through absorption, for example ionization caused by a 
bosonic dark matter particle is analogous to photons ionizing atoms via the photoelectric effect. In each case, the 
rate of interactions depends on the amount of dark matter near the Earth, its velocity and its detailed properties. The 
rate of interactions is also expected to modulate through the year due to the {peculiar} motion of the Earth about the 
Sun, and the Sun about the Galactic center. 

In all of these cases, we need some way of observing the recoiling nuclei or electrons, and of measuring the energy spectrum of the recoiling species, which encodes clues about the dark matter properties. The energy transferred to a 
detector by dark matter interactions can be picked up using three main signals, which depend on which detector 
medium is being used. The first is that target nuclei may be excited, followed by the release of photons when they 
de-excite (so-called scintillation). The second is that heat may be produced, which would cause 
phonon excitations, i.e.~vibrations of the atoms in a crystal. Finally, the interaction might lead directly to the ionisation of atoms within 
the detector. 

Specific detector technologies that exploit one or more of these effects have been developed, and more are being 
proposed all the time. It is particularly useful to build detectors that can see two of the signals, since the relative 
amount of each type of signal (e.g.~scintillation versus heat) strongly depends on the type of particle interacting with 
the detector. This means that one can in principle discriminate nuclear recoil events from electron recoil events, and thus dark matter-induced interactions from backgrounds. 
By suppressing backgrounds, the leading experiments (like LZ, XENON or PandaX) can be sensitive to a single scattering event per year, depositing as little as a few keV of nuclear recoil energy in a ton-scale detector.

\subsection{Dark matter production}

The potential ability of dark matter to interact with electrons or quarks gives us a third way to search for its 
non-gravitational interactions. The Large Hadron Collider (LHC) at CERN -- a particle accelerator at the 
`energy frontier' -- 
smashes protons together at a centre of mass energy of 14 TeV. Since these are made of quarks and gluons, it is 
natural to suppose that any possible interaction between dark matter and quarks would allow us to make dark matter 
at the LHC from scratch. Such a process could be observed by looking for evidence in the detectors that surround the 
collision, whose job it is to measure the energy and momentum of any outgoing particles produced by the collider. 

Dark matter cannot be seen directly by the detectors of the LHC since it does not interact strongly 
enough to leave visible traces. However, all is not lost. Dark matter would almost certainly be produced in conjunction 
with visible matter, either directly or through the decay of a Standard Model particle such as the Z boson or Higgs 
boson. We can therefore look for events in the detectors where the amount of observed momentum  
in the event does not match the amount that we started with. At the LHC, this is complicated by the fact that the 
precise centre of mass energy (which is equivalent to “the amount of momentum we started with”) is not known. Some 
bit of proton hits some other bit of proton, but we cannot say with certainty what occurred in order to apply 
conservation of momentum. What we can do instead is look at 
the components of momentum in the plane transverse to the beam directions, in which we can say with certainty that 
there was no momentum before the collision, and hence that there must be a complete balance of 
momenta for the particles after the collision. Any discrepancy indicates a missing component, which could arise from 
escaping dark matter, but might also result from the production of Standard Model neutrinos. The same logic that applies to the LHC also applies to high energy electron-positron colliders, except that electrons and positrons have no 
substructure and we know the exact center-of-mass energy. This makes the detection of possible dark matter 
signatures even easier. 

Yet another way to search for dark matter production is to study the properties of mesons (i.e.~bound states of quark-antiquark pairs) 
whose decay rates are measured very precisely. If dark matter was light enough to be produced in hadron decays, 
one could observe an anomalous decay rate to final states with missing momentum. Such measurements are best done at 
`intensity frontier' experiments, for example at lower-energy colliders where experiments 
like Belle II study $B$ mesons with exceptional precision. 
Alternatively, one can smash a highly energetic beam of particles into a fixed target that absorbs the beam energy, a setup known as a `beam dump’’ experiment. These experiments are typically much cheaper and easier to perform than colliders operating with two beams, and 
the range of incident beams includes protons, electrons and muons. While the event rate can be much higher than for 
colliders, only a small fraction of the incident beam energy is available for creating new particles, which means that 
beam dump experiments like NA64 or the planned SHiP facility can only search for dark matter that is comprised of 
sufficiently light particles.	
{One can also combine this technology with the techniques discussed in the previous section and attempt to observe 
the scattering of dark matter particles produced in a beam dump experiment in a target placed at some distance from 
the interaction point.}

One important aspect of accelerator searches for dark matter is that they cannot, on their own, reveal that any new 
matter produced is the same as the dark matter that we see interacting gravitationally out in space. One reason is that 
the dark matter might escape the detector in fractions of a second, but could easily decay shortly afterwards without 
anyone knowing, in which case it would not be stable on cosmological timescales. Another reason is that there is no 
way of knowing we have produced the dominant DM component that is observed cosmologically and astrophysically. 
We must therefore combine collider observations with more direct observations in order to properly understand the 
origin of dark matter. 

\subsection{Dark matter conversion}
Certain dark matter candidates are stable (or nearly stable) in vacuum, but can be converted into known particles (and 
vice versa) in suitable environments such as strong magnetic fields or the dense plasma in stellar cores. If such 
conversion processes can happen, they give rise to modified versions of the detection strategies discussed above. 
For example, one can search for dark matter particles produced in astrophysical objects, either by directly detecting 
the produced particles on Earth or by observing the effect that DM production has on emission spectra or on stellar 
evolution. Promising targets are particularly hot and dense stars, such as white dwarfs, stars in special points of their 
evolution, such as red giants or supernovae, or stars with large magnetic fields, such as pulsars. But also the Sun 
may be a source of such dark matter particles, which could be detected with so-called helioscopes like CAST or the 
planned IAXO experiment.

Alternatively, one can attempt to create a suitable environment on Earth to convert particles from the local dark matter 
density into a measurable signal. These haloscope experiments often employ resonant cavities that can be tuned to 
specific frequencies to enhance the conversion probability. The currently leading sensitivities are achieved by 
the ADMX experiment. Finally, so-called light-shining-through-wall experiments attempt to produce dark matter 
particles on Earth using high-intensity lasers in strong magnetic fields. The idea of these experiments is that the 
photons from the laser may be converted into dark matter particles, which can then pass through an otherwise 
intransparent wall in order to be detected on the other side.

\subsection{The cosmos as a dark matter laboratory}

As already mentioned, non-gravitational interactions of dark matter particles with ordinary matter would potentially 
also show up in various cosmological observables. For example, {the comparison of BBN theory with observations 
tells us} that dark matter lighter 
than a few MeV can never have been in thermal equilibrium with the primordial plasma, which in turn
limits the allowed interaction strength with Standard Model 
particles. Even for higher dark matter masses, cosmological observations  put stringent bounds on the scattering of
dark matter with ordinary electrons, protons and neutrinos. In particular, such scattering would affect the evolution of 
density perturbations in both dark matter and baryon components in a way that leaves visible imprints on the CMB
temperature power spectrum.

Intriguingly, cosmological probes are not restricted to interactions that in principle could also be tested in experiments 
on Earth. Instead, they offer additional, fully complementary tests that even apply
to dark matter particles interacting so feebly with ordinary matter that they would escape detection 
even in hypothetical high-performance  detectors. We have already discussed in some detail one type of such 
interactions, namely dark matter self-scattering (cf.~figure \ref{fig:BC}). 
We have also mentioned that free-streaming, i.e.~dark matter particles moving `too fast', 
leads to a suppression of small-scale structure. Importantly, this general requirement holds independently
of whether the dark matter particles' large velocity is caused by interactions with ordinary matter or by some
other mechanism. In fact, the exact shape of such a suppression in the spectrum of matter density perturbations  
can carry valuable information about dark matter interactions not visible otherwise.
A third example in this category is dark matter decaying, or annihilating, 
into new light particles. Even if both dark matter and these new particles are otherwise completely invisible, 
such processes would still correspond to a conversion from non-relativistic matter to relativistic radiation. 
The fact that such energy densities, and their perturbations, evolve differently 
as the universe expands can be used to infer that at most $\sim10\%$ of dark matter can have been converted
in this way since CMB times. Notably, this conclusion is completely independent of the dark matter candidate and 
conversion mechanism.

There is a plethora of upcoming and already ongoing missions, such as DESI, the Vera C.\ Rubin Observatory and 
the EUCLID satellite,  that will probe dark matter interactions
even in completely secluded dark sectors, i.e.~without necessarily having any appreciable coupling to ordinary 
matter. These experiments range from large-scale galaxy surveys to detailed probes of the only mildly non-linear 
power spectrum (based on observing atomic transitions of neutral hydrogen in the early universe that emit photons 
with a characteristic wavelength of 21\,cm).
Upcoming searches for cosmological gravitational wave backgrounds, for example the planned LISA mission,
can also help to constrain certain dark matter 
models.

\section{Dark matter candidates}
\label{sec:models}

When we speculate about the particle nature of dark matter, we should first consider the possibility that the known 
elementary particles
mimic the observed properties of dark matter.  After all, the Standard Model of particle physics is exceptionally 
successful in explaining a wide range of phenomena.  
However, any electrically charged  particles -- such as quarks and charged leptons 
-- directly interact with photons, which excludes them from being dark matter.  
Non-stable particles can also immediately be discarded as dark matter candidates.

Of the known fundamental particles, only neutrinos are electrically neutral and stable. 
However, experimental bounds on the tiny neutrino mass imply that these particles would still be 
moving at relativistic speeds during the early stages of structure formation, preventing them from 
collapsing into overdensities. 
Moreover, 
as mentioned above, Standard Model neutrinos strongly violate the Tremaine-Gunn bound on the mass of fermionic 
dark matter particles from the Pauli exclusion principle. There have also been various proposals to explain dark matter 
in terms of bound states or condensates of the known fundamental particles, but none of these enjoy universal success.  

Most scientists therefore assume that dark matter is made of one or more new, yet undiscovered elementary 
particles.  While at face value this sounds like a bold assumption, surprisingly many extensions of the Standard Model 
contain particles with just the right properties. 
Intriguingly, many of these new particles were 
originally introduced to solve problems of the Standard Model entirely independent from the dark matter puzzle. 
Here we will not 
review these theories in detail but instead focus on a few representative examples, highlighting typical considerations
when identifying well-motivated and viable dark matter candidates.

\subsection{Case study: Heavy neutrinos}

Since Standard Model neutrinos are excluded as dark matter candidates because of their small mass, an obvious way 
to construct a potentially viable dark matter model would be to postulate a hypothetical counterpart of ordinary 
neutrinos, which interacts in the same way but has a substantially larger mass $M_\nu\gg$\,eV.  Such heavy neutrinos 
naturally arise in various extensions of the Standard Model, including those that attempt to explain the origin of 
neutrino mass. Indeed, historically, heavy neutrinos were one of the 
earliest candidates proposed for dark matter. 

Once the free-streaming bound and the Tremaine-Gunn bound are satisfied, the remaining key question is whether 
heavy neutrinos can be produced in the early universe with sufficient abundance to account for the cosmologically 
observed dark matter density. Indeed, the evolution of their number density in the early universe is governed by a 
process known as thermal freeze-out, based on well-understood principles from thermodynamics. 
The basic 
idea  is that the interactions of heavy neutrinos with other Standard Model particles were sufficient 
to establish thermal equilibrium in the hot and dense early universe, with creation and annihilation processes
of these particles exactly balancing each other. As the universe expanded and cooled down, 
the rate of these interactions decreased and became less and less relevant. Eventually, the number density of the 
heavy neutrinos only changed with the volume of the expanding universe, completely unaffected
by any microscopic interactions.

Performing these calculations in detail, 
one finds that for sufficiently large dark matter masses the particles thus produced are, as required, highly 
non-relativistic at the onset of structure formation. 
The abundance, furthermore, turns out to be largely independent of the dark matter mass in this case, 
but inversely proportional to 
the thermally averaged annihilation cross-section $\langle \sigma v \rangle$
of heavy neutrinos (and anti-neutrinos) to ordinary particles. 
This result makes intuitive sense, because larger annihilation cross sections keep the dark matter particles
in equilibrium for longer, i.e.~until the universe has cooled down to smaller temperatures -- and thermodynamics tells 
us that the abundance of very massive particles in a plasma is highly suppressed at small temperatures.
Comparing the theoretical prediction to the observed dark matter density allows us to accurately determine the 
thermally averaged annihilation cross section as\footnote{%
The quantity $\langle \sigma v \rangle$ is the most important  parameter in the context of 
dark matter produced via the freeze-out mechanism, but also in the context of indirect searches for such candidates.
Commonly abbreviated as the `thermally averaged annihilation cross section' for simplicity, 
it refers to the velocity-weighted annihilation cross section, averaged over all dark matter velocities. 
When multiplied by 
the dark matter number density, it describes the average rate at which annihilation processes happen.
} 
\begin{equation}
    \label{eq:thermalsv}
    \langle \sigma v \rangle \approx 6 \times 10^{-26} \, \mathrm{cm^3 \, s^{-1}} \; .
\end{equation}
This result is of central importance for dark matter physics because it provides guidance for the construction of viable 
models far beyond the specific example presented here. 

Returning to the heavy neutrino, it is instructive to consider two limiting cases. 
For $M_\nu$ smaller than the $Z$ boson mass, $M_Z$, one obtains
\begin{equation}
  \langle \sigma v \rangle \sim \frac{g^4 M_\nu^2}{M_Z^4} \,,
\label{eq:sigmav_light}
\end{equation}
where $g \approx 0.3$ is the coupling strength of the weak interaction. In this limit, a smaller dark matter mass thus
leads to a smaller cross section and hence a larger dark matter abundance.
Quantitatively, the requirement that the abundance of heavy neutrinos does not exceed the total amount of dark 
matter implies $M_\nu \gtrsim 2$ GeV, known as the Lee-Weinberg bound.
For $M_\nu \gtrsim M_Z$, on the other hand, one finds
\begin{equation}
  \langle \sigma v \rangle \sim \frac{g^4}{M_\nu^2} \,,
\label{eq:sigmav_heavy}
\end{equation}
which numerically reproduces eq.~(\ref{eq:thermalsv}), and hence the measured value of the dark matter abundance,
for $M_\nu \sim 1$ TeV. 

While theoretically very appealing, heavy neutrino properties are stringently constrained by 
astrophysical observations, cosmological measurements, and laboratory experiments.
In particular, for $M_\nu < M_Z/2$, the existence of additional neutrinos is excluded by searches for unobserved,
so-called invisible decays of the $Z$ boson. 
Heavier masses, on the other hand, are strongly constrained by searches for dark matter scattering.
In combination, these constraints strongly disfavour heavy versions of ordinary neutrinos as dark matter candidates.

\subsection{General thermal relics}
 
 The discussion of heavy neutrinos provides an intriguing historical example of how combined experimental efforts
 on a global scale allowed physicists to test, and eventually rule out, an {\it a priori} extremely well-motivated dark matter 
 candidate. It also provides  very useful guidance for the construction of viable dark matter models in general.
 A logical next step is to consider other scenarios where the dark matter particle is  
 somehow connected to the weak interactions of the Standard Model (or the Higgs mechanism, being 
 tightly intertwined with these interactions). Indeed, theoretical considerations completely independent of 
 dark matter have been put forward in the past to argue that new particles and other effects of new physics 
 should become visible at TeV energies, i.e.~at energies only slightly higher than the mass of the heaviest known 
 fundamental particles such as the Higgs boson and the top quark. Dark matter particles that appear in such theories, and are 
 produced via thermal freeze-out, are generally referred to as Weakly Interacting Massive 
Particles (WIMPs).  
The seeming coincidence that new physics effects were expected to appear at the TeV scale and
that new neutral and stable particles at this scale would almost automatically be produced in the cosmologically 
required amount to account for all of the observed dark matter
-- as discussed in the context of eq.~(\ref{eq:sigmav_heavy}) --   has historically been dubbed the `WIMP miracle'.

WIMPs are constrained by similar types of observations as the heavy neutrino. In fact, the theoretical emphasis on the 
WIMP dark matter paradigm has been a significant driver for the experimental program described in 
 section~\ref{sec:searches}. This 
 implies, in particular, that any dark matter candidate that primarily interacts via $Z$ boson exchange is excluded.
Even beyond this, the plethora of ongoing searches has started to put the WIMP hypothesis under significant 
pressure, though well-motivated islands of viable parameter space remain that can be tested with upcoming
searches. A prominent example of the latter are TeV-scale dark matter particles
interacting dominantly via Higgs-boson exchange, such as the `partner'  of the Higgs boson in theories of supersymmetry,
the so-called Higgsino.

We can generalize the lessons learned from the heavy neutrino example in yet another direction, namely
by realizing that thermal production via freeze-out does in fact not require weak-scale interactions.
This is because the condition for eq.~\eqref{eq:sigmav_heavy} to match eq.~\eqref{eq:thermalsv}
only fixes the coupling to mass {\it ratio},  not any of these quantities individually.
One could, in other words, postulate the existence of a new interaction that is responsible for keeping
dark and ordinary matter in thermal equilibrium in the early universe, where the coupling $g$ is now a free parameter
of the new theory. Increasing $g$ then makes it possible to also increase the allowed dark 
  matter mass, potentially evading experimental constraints. However, on theoretical grounds one requires 
  $g < \sqrt{4\pi}$, which implies a bound on the dark matter mass of roughly $M_{\rm DM} \lesssim 100$ TeV, 
  known as the unitarity bound.\footnote{Recent research,
however, suggests that the idea of a strict upper mass limit based on unitarity may be misleading, and that there is no 
clear, model-independent cap on the mass of thermally produced dark matter.  This challenges some previous 
assumptions and opens the door to exploring heavier dark matter candidates.}

Introducing a new interaction would not only modify the value of $g$, but necessarily also replace the $Z$ boson
with another, hypothetical  force carrier $Z'$ that mediates the new interaction. 
{This opens up an avenue to evade the Lee-Weinberg bound and construct dark matter models with a mass below the 
GeV scale. By making $g$ and/or $M_{Z'}$ sufficiently small, it is possible to reproduce the annihilation cross section 
in eq.~\eqref{eq:thermalsv} both for $M_\text{DM} \geq M_{Z'}$, see eq.~\eqref{eq:sigmav_heavy}, and for 
$M_\text{DM} \leq M_{Z'}$, see eq.~\eqref{eq:sigmav_light}. In both cases, the new mediator must have a mass much 
smaller than the $Z$ boson mass, which means that it can be independently searched for experimentally.}
However, to ensure agreement with cosmological data, freeze-out must always happen before primordial element 
formation during BBN, as discussed in section \ref{chap1:sec1}, which implies a rather model-independent lower 
bound on the dark matter mass of a few MeV.

 \begin{figure}[t]
\centering
\includegraphics[width=.85\textwidth]{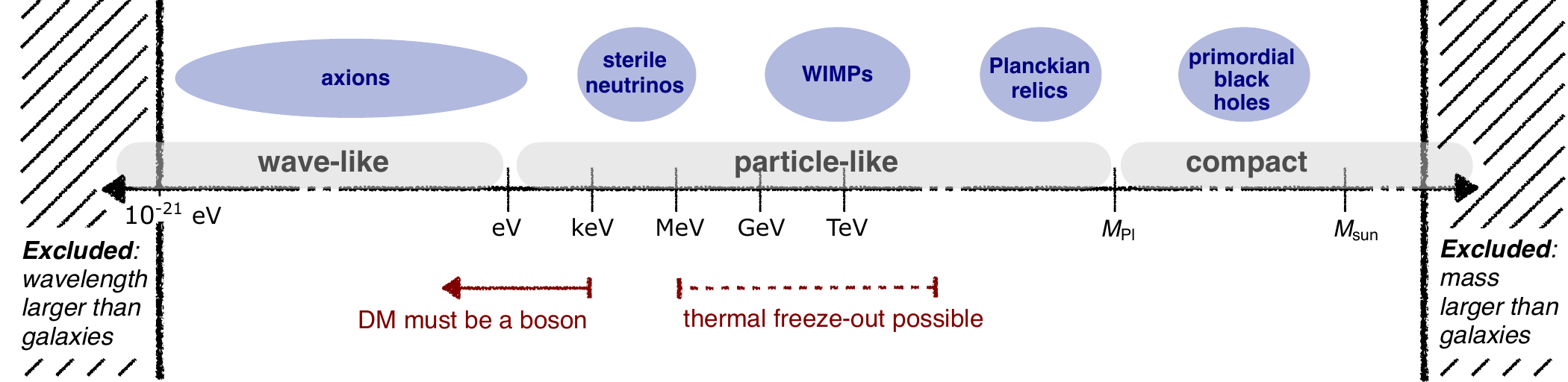}
\vspace{1.5mm}
\caption{Possible mass range of dark matter particles, along with a few selected candidates mentioned in the text. 
Importantly, the description in terms of elementary particles necessarily breaks down both for very large and for very 
small masses. For masses exceeding the Planck mass, $M_{\rm Pl}\sim10^{19}\,{\rm GeV}\sim10^{-5}\,{\rm g}$, dark 
matter rather consists of compact, macroscopic objects. For masses below about $1$~eV, on the other hand, dark 
matter is better described by collective, wave-like phenomena. Intuitively, this is because
the measured overall energy density in dark matter implies that the typical separation 
between two `particles' will be smaller than their De-Broglie  wavelength in that case.
}
\label{fig:DM_mass}
\end{figure}

\subsection{Non-thermal relics}

All the arguments above rely on the assumption of thermal freeze-out. If we relax this assumption, many exciting new 
directions open up. A well-studied idea is that of so-called sterile (or right-handed) neutrinos, which do not directly interact 
with $Z$ bosons, but can be converted into ordinary (active) neutrinos with a tiny probability. Likewise, any known 
process that produces ordinary neutrinos has a tiny probability to instead produce a sterile neutrino. Such
processes are so rare that sterile neutrinos 
can never enter into thermal equilibrium. Rather than having their relic abundance set via the freeze-out 
mechanism, such particles would be produced via a process called freeze-in, which can be thought of as a gradual 
leakage of energy from the thermal bath of Standard Model particles into the `dark sector'.

Because they hardly interact at all, sterile neutrinos would evade all laboratory searches for dark matter. Moreover, they 
do not affect the primordial element abundances and therefore can have a mass below the MeV scale. Nevertheless, 
sterile neutrinos are predicted to decay into known neutrinos on cosmological timescales.
If such a decay is accompanied by the emission of a photon, one obtains a potentially observable X-ray signal. 
Indeed, searches for such X-ray emission place strong constraints on models of sterile neutrinos.

Sterile neutrinos still have to satisfy the warm dark matter bound from structure formation and the Tremaine-Gunn 
bound from the Pauli exclusion principle, which requires their mass to be heavier than a few keV. Both bounds can be 
evaded in models of wave-like dark matter. In these models, dark matter is assumed to be a bosonic particle with such 
a small mass ($m \lesssim 1$ eV) that huge occupation numbers are needed to reproduce the observed dark matter 
abundance. This leads to classicalisation: Large collections of particles can be equivalently described by classical 
fields that oscillate with a frequency $\omega$ given by the rest mass of the particles. 
In this picture of wave-like dark matter, the requirement that dark matter is cold translates to the requirement that the 
oscillations are coherent, i.e.~that the phase is approximately constant over {sufficiently large distances}. 
Such coherent 
oscillations can be generated in the early universe via the so-called realignment mechanism, which assumes that the 
field is initially displaced from the minimum of its potential by a certain offset. 
The dark matter density is then proportional to the squared amplitude of these oscillations.

The most famous example for wave-like dark matter are so-called axions, which were originally proposed to explain the unexpected smallness of the neutron electric dipole moment (referred to for historical reasons as the `strong CP problem'). Given 
experimental and astrophysical constraints, they are predicted to have mass below 1 meV and extremely weak 
interactions with ordinary matter. Nevertheless, their collective effect, i.e.~the coherent field oscillations, may lead to 
distinct observable effects, such as oscillating dipole moments or oscillating fundamental constants that can be picked 
up with quantum sensors such as atomic clocks or interferometers.

At the opposite end of the mass scale, extremely heavy dark matter particles could also be produced at the 
earliest stages of the cosmological evolution, at the end of an era with exponential expansion referred to as inflation.
Such `Planckian relics', which could be as heavy as the Planck mass, $M_{\rm Pl}\sim10^{19}\,{\rm GeV}$, 
might interact only gravitationally, and hence be extremely challenging to detect experimentally.
It is even conceivable that dark matter consists of still more massive objects.
In this case, the (fundamental) particle description based on quantum mechanics
breaks down, and one should rather think of classical, compact objects. An extreme example in this latter category
are primordial black holes, i.e.~black holes that were formed in the early universe. Since their existence predates star 
formation, they cannot originate from the collapse of astrophysical objects and require a new, unknown production 
mechanism.

Let us end by stressing that the above list of potential particle dark matter candidates is by no means exhaustive. 
Recent years have seen a boost of activity in identifying further alternatives, both in a theory-driven manner (based on 
guiding principles similar to those outlined above) and by a more phenomenological approach to explore 
what type of candidates could in principle 
be probed by upcoming experimental searches. As the remaining parameter space for classical WIMP-like candidates
keeps shrinking, the focus of these efforts has somewhat shifted towards lighter particles and in particular non-thermal relics.

\section{Summary}
\label{sec:summary}

It is impossible to describe the universe at galactic and larger scales with known physics alone, i.e.~within the 
framework provided by the Standard Model of particle physics and general relativity.
When defined as the missing ingredient to describe cosmological structure formation, it is thus clear that
dark matter has already been observed. What is truly remarkable in this perspective is that an extremely large range 
of independent cosmological
observations related to structure formation can be described by just one single phenomenological parameter,
namely the average energy density of dark matter present in the universe today. In the context of the cosmological 
standard model, in fact, this parameter can be measured with an accuracy reaching the sub-percent level.

In a sense, this situation sets the gold standard: any alternative explanation of the dark matter phenomenon
would have to first get the basic cosmological picture right. Importantly, this is most relevant for the largest observable
scales, such as those relevant for observations of the CMB or large-scale galaxy correlation functions, 
because they are mostly based  on linear density perturbations that are very well
understood. Dark matter also manifests itself at smaller scales, in virialized objects like galaxies, but there are
many complicated details in the transition from linear to non-linear evolution that are not yet completely understood,
in particular concerning the impact of ordinary matter on the late stages of structure formation.
This generically makes an unbiased,  direct comparison between theory and observations very challenging at galactic 
scales.

Many modern cosmological observations can be understood in terms of the properties of 
fundamental particles, such as protons, electrons and neutrinos. This motivates attempts to also address the dark 
matter puzzle in terms of new elementary particles. While even very basic properties of such a dark matter particle
currently remain unknown  -- such as its mass, cf.~figure \ref{fig:DM_mass} -- it is worth stressing that there exist 
strong and independent 
motivations for many of the proposed candidates. Notably, such arguments are often accompanied by clear theoretical predictions
for additional interactions that can be used to probe these candidates either directly in laboratory experiments or
indirectly due to their impact on cosmological and astrophysical observables.

There is a massive ongoing experimental program to search for dark matter, in the sense of directly confirming 
its particle identity, that is motivated by the theoretical considerations sketched above. In fact, the scale of these global 
efforts adequately reflects the scientific importance of this quest, the identity of dark matter regularly 
being ranked among the top 10 open fundamental science questions. 
A paradigm shift away from the traditional theoretical focus on WIMP-like dark matter has only recently 
started to trigger a corresponding broadening of experimental activities and searches that are targeted
at detecting other classes of dark matter candidates. 
The expected progress both in these new directions and in testing the remaining islands of still viable WIMP
models implies particularly exciting times ahead. In fact, it is fully conceivable that the new parts of 
parameter and model space tested by the next generation of instruments will contain the answer to
an almost century-old problem.

\begin{ack}[Acknowledgments]

C.B. is supported by the Australian Research Council grants DP210101636, DP220100643 and LE21010001. TB and 
FK are grateful to the Mainz Institute for Theoretical Physics (MITP) of the Cluster of Excellence PRISMA+ 
(Project ID 390831469), for its hospitality and its partial support during the completion of this work. MW is supported 
by the ARC Centre of Excellence for Dark Matter Particle Physics CE20010000.
We thank Morten Tryti Berg, Alexander Eberhart and Edvard R\o rnes for valuable feedback on the manuscript.
\end{ack}

\seealso{
Further details can be found in the dark matter lectures by \cite{ Gelmini:2015zpa,Lisanti:2016jxe, Safdi:2022xkm}, as well as in the following more specialized lectures and reviews:
\begin{itemize}
    \item \textbf{Dark matter in cosmology and astrophysics:} \cite{Cline:2018fuq} on the role of dark matter in cosmology, \cite{Frenk:2012ph} on structure formation, \cite{Read:2014qva} on the local dark matter density
    \item \textbf{Dark matter searches:} \cite{Bringmann:2012ez,Gaskins:2016cha,Slatyer:2021qgc} on searches for dark matter annihilation or decay, \cite{Cerdeno:2010jj,Lin:2019uvt} on searches for dark matter scattering, \cite{Kahlhoefer:2017dnp} on collider searches for dark matter, 
    \cite{Graham:2015ouw} on searches for dark matter conversion
    \item \textbf{Specific dark matter candidates:} \cite{Feng:2010gw} on thermal dark matter, 
    \cite{Abazajian:2017tcc} on sterile neutrinos,\cite{Marsh:2015xka} on axions,  \cite{Green:2020jor} on primordial black holes
\end{itemize}
}

\begin{thebibliography*}{23}
\providecommand{\bibtype}[1]{}
\providecommand{\natexlab}[1]{#1}
{\catcode`\|=0\catcode`\#=12\catcode`\@=11\catcode`\\=12
|immediate|write|@auxout{\expandafter\ifx\csname
  natexlab\endcsname\relax\gdef\natexlab#1{#1}\fi}}
\renewcommand{\url}[1]{{\tt #1}}
\providecommand{\urlprefix}{URL }
\expandafter\ifx\csname urlstyle\endcsname\relax
  \providecommand{\doi}[1]{doi:\discretionary{}{}{}#1}\else
  \providecommand{\doi}{doi:\discretionary{}{}{}\begingroup
  \urlstyle{rm}\Url}\fi
\providecommand{\bibinfo}[2]{#2}
\providecommand{\eprint}[2][]{\url{#2}}

\bibtype{Article}%
\bibitem[Abazajian(2017)]{Abazajian:2017tcc}
\bibinfo{author}{Abazajian KN} (\bibinfo{year}{2017}).
\bibinfo{title}{{Sterile neutrinos in cosmology}}.
\bibinfo{journal}{{\em Phys. Rept.}} \bibinfo{volume}{711-712}:
  \bibinfo{pages}{1--28}. \bibinfo{doi}{\doi{10.1016/j.physrep.2017.10.003}}.
\eprint{1705.01837}.

\bibtype{Book}%
\bibitem[Bauer and Plehn(2019)]{Bauer:2017qwy}
\bibinfo{author}{Bauer M} and  \bibinfo{author}{Plehn T}
  (\bibinfo{year}{2019}).
\bibinfo{title}{{Yet Another Introduction to Dark Matter}: {The Particle
  Physics Approach}}, \bibinfo{series}{Lecture Notes in Physics},
  \bibinfo{volume}{959}, \bibinfo{publisher}{Springer}.
\bibinfo{doi}{\doi{10.1007/978-3-030-16234-4}}.
\eprint{1705.01987}.

\bibtype{Article}%
\bibitem[Bertone and Hooper(2018)]{Bertone:2016nfn}
\bibinfo{author}{Bertone G} and  \bibinfo{author}{Hooper D}
  (\bibinfo{year}{2018}).
\bibinfo{title}{{History of dark matter}}.
\bibinfo{journal}{{\em Rev. Mod. Phys.}} \bibinfo{volume}{90}
  (\bibinfo{number}{4}): \bibinfo{pages}{045002}.
  \bibinfo{doi}{\doi{10.1103/RevModPhys.90.045002}}.
\eprint{1605.04909}.

\bibtype{Article}%
\bibitem[Bringmann and Weniger(2012)]{Bringmann:2012ez}
\bibinfo{author}{Bringmann T} and  \bibinfo{author}{Weniger C}
  (\bibinfo{year}{2012}).
\bibinfo{title}{{Gamma Ray Signals from Dark Matter: Concepts, Status and
  Prospects}}.
\bibinfo{journal}{{\em Phys. Dark Univ.}} \bibinfo{volume}{1}:
  \bibinfo{pages}{194--217}. \bibinfo{doi}{\doi{10.1016/j.dark.2012.10.005}}.
\eprint{1208.5481}.

\bibtype{Article}%
\bibitem[Cerdeno and Green(2010)]{Cerdeno:2010jj}
\bibinfo{author}{Cerdeno DG} and  \bibinfo{author}{Green AM}
  (\bibinfo{year}{2010}), \bibinfo{month}{2}.
\bibinfo{title}{{Direct detection of WIMPs}} :
  \bibinfo{pages}{347--369}\bibinfo{doi}{\doi{10.1017/CBO9780511770739.018}}.
\eprint{1002.1912}.

\bibtype{Article}%
\bibitem[Cirelli et al.(2024)]{Cirelli:2024ssz}
\bibinfo{author}{Cirelli M}, \bibinfo{author}{Strumia A} and
  \bibinfo{author}{Zupan J} (\bibinfo{year}{2024}), \bibinfo{month}{6}.
\bibinfo{title}{{Dark Matter}} \eprint{2406.01705}.

\bibtype{Article}%
\bibitem[Cline(2019)]{Cline:2018fuq}
\bibinfo{author}{Cline JM} (\bibinfo{year}{2019}).
\bibinfo{title}{{TASI Lectures on Early Universe Cosmology: Inflation,
  Baryogenesis and Dark Matter}}.
\bibinfo{journal}{{\em PoS}} \bibinfo{volume}{TASI2018}: \bibinfo{pages}{001}.
\eprint{1807.08749}.

\bibtype{Article}%
\bibitem[Feng(2010)]{Feng:2010gw}
\bibinfo{author}{Feng JL} (\bibinfo{year}{2010}).
\bibinfo{title}{{Dark Matter Candidates from Particle Physics and Methods of
  Detection}}.
\bibinfo{journal}{{\em Ann. Rev. Astron. Astrophys.}} \bibinfo{volume}{48}:
  \bibinfo{pages}{495--545}.
  \bibinfo{doi}{\doi{10.1146/annurev-astro-082708-101659}}.
\eprint{1003.0904}.

\bibtype{Article}%
\bibitem[Frenk and White(2012)]{Frenk:2012ph}
\bibinfo{author}{Frenk CS} and  \bibinfo{author}{White SDM}
  (\bibinfo{year}{2012}).
\bibinfo{title}{{Dark matter and cosmic structure}}.
\bibinfo{journal}{{\em Annalen Phys.}} \bibinfo{volume}{524}:
  \bibinfo{pages}{507--534}. \bibinfo{doi}{\doi{10.1002/andp.201200212}}.
\eprint{1210.0544}.

\bibtype{Article}%
\bibitem[Gaskins(2016)]{Gaskins:2016cha}
\bibinfo{author}{Gaskins JM} (\bibinfo{year}{2016}).
\bibinfo{title}{{A review of indirect searches for particle dark matter}}.
\bibinfo{journal}{{\em Contemp. Phys.}} \bibinfo{volume}{57}
  (\bibinfo{number}{4}): \bibinfo{pages}{496--525}.
  \bibinfo{doi}{\doi{10.1080/00107514.2016.1175160}}.
\eprint{1604.00014}.

\bibtype{Inproceedings}%
\bibitem[Gelmini(2015)]{Gelmini:2015zpa}
\bibinfo{author}{Gelmini GB} (\bibinfo{year}{2015}), \bibinfo{title}{{The hunt
  for dark matter.}}, \bibinfo{booktitle}{{Theoretical Advanced Study Institute
  in Elementary Particle Physics}: {Journeys Through the Precision Frontier:
  Amplitudes for Colliders}},  \bibinfo{pages}{559--616}, \eprint{1502.01320}.

\bibtype{Article}%
\bibitem[Graham et al.(2015)]{Graham:2015ouw}
\bibinfo{author}{Graham PW}, \bibinfo{author}{Irastorza IG},
  \bibinfo{author}{Lamoreaux SK}, \bibinfo{author}{Lindner A} and
  \bibinfo{author}{van Bibber KA} (\bibinfo{year}{2015}).
\bibinfo{title}{{Experimental Searches for the Axion and Axion-Like
  Particles}}.
\bibinfo{journal}{{\em Ann. Rev. Nucl. Part. Sci.}} \bibinfo{volume}{65}:
  \bibinfo{pages}{485--514}.
  \bibinfo{doi}{\doi{10.1146/annurev-nucl-102014-022120}}.
\eprint{1602.00039}.

\bibtype{Article}%
\bibitem[Green and Kavanagh(2021)]{Green:2020jor}
\bibinfo{author}{Green AM} and  \bibinfo{author}{Kavanagh BJ}
  (\bibinfo{year}{2021}).
\bibinfo{title}{{Primordial Black Holes as a dark matter candidate}}.
\bibinfo{journal}{{\em J. Phys. G}} \bibinfo{volume}{48} (\bibinfo{number}{4}):
  \bibinfo{pages}{043001}. \bibinfo{doi}{\doi{10.1088/1361-6471/abc534}}.
\eprint{2007.10722}.

\bibtype{Article}%
\bibitem[Kahlhoefer(2017)]{Kahlhoefer:2017dnp}
\bibinfo{author}{Kahlhoefer F} (\bibinfo{year}{2017}).
\bibinfo{title}{{Review of LHC Dark Matter Searches}}.
\bibinfo{journal}{{\em Int. J. Mod. Phys. A}} \bibinfo{volume}{32}
  (\bibinfo{number}{13}): \bibinfo{pages}{1730006}.
  \bibinfo{doi}{\doi{10.1142/S0217751X1730006X}}.
\eprint{1702.02430}.

\bibtype{Article}%
\bibitem[Lin(2019)]{Lin:2019uvt}
\bibinfo{author}{Lin T} (\bibinfo{year}{2019}).
\bibinfo{title}{{Dark matter models and direct detection}}.
\bibinfo{journal}{{\em PoS}} \bibinfo{volume}{333}: \bibinfo{pages}{009}.
  \bibinfo{doi}{\doi{10.22323/1.333.0009}}.
\eprint{1904.07915}.

\bibtype{Inproceedings}%
\bibitem[Lisanti(2017)]{Lisanti:2016jxe}
\bibinfo{author}{Lisanti M} (\bibinfo{year}{2017}), \bibinfo{title}{{Lectures
  on Dark Matter Physics}}, \bibinfo{booktitle}{{Theoretical Advanced Study
  Institute in Elementary Particle Physics}: {New Frontiers in Fields and
  Strings}},  \bibinfo{pages}{399--446}, \eprint{1603.03797}.

\bibtype{Book}%
\bibitem[Mambrini(2021)]{Mambrini:2021cwd}
\bibinfo{author}{Mambrini Y} (\bibinfo{year}{2021}).
\bibinfo{title}{{Particles in the Dark Universe. A Student\textquoteright{}s
  Guide to Particle Physics and Cosmology}}, \bibinfo{publisher}{Springer}.
\bibinfo{comment}{ISBN} \bibinfo{isbn}{978-3-030-78138-5, 978-3-030-78139-2}.
\bibinfo{doi}{\doi{10.1007/978-3-030-78139-2}}.

\bibtype{Article}%
\bibitem[Marsh(2016)]{Marsh:2015xka}
\bibinfo{author}{Marsh DJE} (\bibinfo{year}{2016}).
\bibinfo{title}{{Axion Cosmology}}.
\bibinfo{journal}{{\em Phys. Rept.}} \bibinfo{volume}{643}:
  \bibinfo{pages}{1--79}. \bibinfo{doi}{\doi{10.1016/j.physrep.2016.06.005}}.
\eprint{1510.07633}.

\bibtype{Book}%
\bibitem[Marsh et al.(2024)]{Marsh:2024ury}
\bibinfo{author}{Marsh DJE}, \bibinfo{author}{Ellis D} and
  \bibinfo{author}{Mehta VM} (\bibinfo{year}{2024}), \bibinfo{month}{9}.
\bibinfo{title}{{Dark Matter: Evidence, Theory, and Constraints}},
  \bibinfo{publisher}{Princeton University Press}.
\bibinfo{comment}{ISBN} \bibinfo{isbn}{978-0-691-24952-0}.

\bibtype{Book}%
\bibitem[Profumo(2017)]{Profumo:2017hqp}
\bibinfo{author}{Profumo S} (\bibinfo{year}{2017}).
\bibinfo{title}{{An Introduction to Particle Dark Matter}},
  \bibinfo{publisher}{World Scientific}.
\bibinfo{comment}{ISBN} \bibinfo{isbn}{978-1-78634-000-9, 978-1-78634-001-6,
  978-1-78634-001-6}.
\bibinfo{doi}{\doi{10.1142/q0001}}.

\bibtype{Article}%
\bibitem[Read(2014)]{Read:2014qva}
\bibinfo{author}{Read JI} (\bibinfo{year}{2014}).
\bibinfo{title}{{The Local Dark Matter Density}}.
\bibinfo{journal}{{\em J. Phys. G}} \bibinfo{volume}{41}:
  \bibinfo{pages}{063101}. \bibinfo{doi}{\doi{10.1088/0954-3899/41/6/063101}}.
\eprint{1404.1938}.

\bibtype{Article}%
\bibitem[Safdi(2024)]{Safdi:2022xkm}
\bibinfo{author}{Safdi BR} (\bibinfo{year}{2024}).
\bibinfo{title}{{TASI Lectures on the Particle Physics and Astrophysics of Dark
  Matter}}.
\bibinfo{journal}{{\em PoS}} \bibinfo{volume}{TASI2022}: \bibinfo{pages}{009}.
  \bibinfo{doi}{\doi{10.22323/1.439.0009}}.
\eprint{2303.02169}.

\bibtype{Article}%
\bibitem[Slatyer(2022)]{Slatyer:2021qgc}
\bibinfo{author}{Slatyer TR} (\bibinfo{year}{2022}).
\bibinfo{title}{{Les Houches Lectures on Indirect Detection of Dark Matter}}.
\bibinfo{journal}{{\em SciPost Phys. Lect. Notes}} \bibinfo{volume}{53}:
  \bibinfo{pages}{1}. \bibinfo{doi}{\doi{10.21468/SciPostPhysLectNotes.53}}.
\eprint{2109.02696}.

\end{thebibliography*}

\end{document}